\documentclass[reprint,10pt,aps,prd,twocolumn,floats,floatfix,showpacs,superscriptaddress,nofootinbib]{revtex4-2}

\usepackage{amssymb,amsmath,amsfonts,amsthm,bm,graphicx}
\usepackage{multirow,txfonts}
\usepackage[usenames]{color}
\usepackage[dvipsnames]{xcolor}
\usepackage[english]{babel}
\usepackage[utf8]{inputenc}
\usepackage[percent]{overpic}
\usepackage[normalem]{ulem}

\xdefinecolor{mylinkcolor}{rgb}{0,0,0.5}
\usepackage[
    bookmarksnumbered, bookmarksopen, bookmarksopenlevel=2,
    breaklinks=true,
    colorlinks=true, filecolor=mylinkcolor, citecolor=mylinkcolor,
    linkcolor=mylinkcolor, urlcolor=mylinkcolor, menucolor=mylinkcolor,
]{hyperref}
\usepackage{cleveref}

\allowdisplaybreaks

\graphicspath{ {figs/} }

\begin{document}

\title{Effective-action model for dynamical scalarization beyond the adiabatic approximation}

\author{Mohammed Khalil}
\email{mohammed.khalil@aei.mpg.de}
\affiliation{Max Planck Institute for Gravitational Physics (Albert Einstein Institute),
Am M\"uhlenberg 1, Potsdam 14476, Germany}
\affiliation{Department of Physics, University of Maryland, College Park, MD 20742, USA}

\author{Raissa F.\ P.\ Mendes}
\email{rfpmendes@id.uff.br}
\affiliation{Instituto de F\'isica, Universidade Federal Fluminense, Niter\'oi, Rio de Janeiro, 24210-346, Brazil.}

\author{N\'{e}stor Ortiz}
\email{nestor.ortiz@nucleares.unam.mx}
\affiliation{Instituto de Ciencias Nucleares, Universidad Nacional Aut\'onoma de M\'exico, Circuito Exterior C.U., A.P. 70-543, M\'exico D.F. 04510, M\'exico.}

\author{Jan Steinhoff}
\email{jan.steinhoff@aei.mpg.de}
\affiliation{Max Planck Institute for Gravitational Physics (Albert Einstein Institute),
Am M\"uhlenberg 1, Potsdam 14476, Germany}

\begin{abstract}
In certain scalar-field extensions to general relativity, scalar charges can develop on compact objects in an inspiraling binary---an effect known as dynamical scalarization.
This effect can be modeled using effective-field-theory methods applied to the binary within the post-Newtonian approximation.
Past analytic investigations focused on the adiabatic (or quasi-stationary) case for quasi-circular orbits.
In this work, we explore the full dynamical evolution around the phase transition to the scalarized regime.
This allows for generic (eccentric) orbits and to quantify nonadiabatic (e.g., oscillatory) behavior during the phase transition.
We also find that even in the circular-orbit case, the onset of scalarization can only be predicted reliably when taking the full dynamics into account, i.e., the adiabatic approximation is not appropriate.
Our results pave the way for accurate post-Newtonian predictions for dynamical scalarization effects in gravitational waves from compact binaries. 
\end{abstract}

\maketitle

\section{Introduction}
\label{sec:intro}

Compact astrophysical objects such as black holes and neutron stars offer special environments for tests of general relativity.
Not only do perturbative corrections to Newtonian gravity become more pronounced, but gravity might also hold surprises in the strong-field regime, due to the possible onset of nonperturbative effects.
A paradigmatic example is the so-called spontaneous scalarization of neutron stars.
In the 90's, Damour and Esposito-Farèse (DEF) showed that in a broad class of scalar extensions of general relativity, the scalar field may remain dormant in the weak-field regime, only to be activated around sufficiently compact material bodies such as neutron stars \cite{Damour:1993hw}.
Although most of the parameter space for spontaneous scalarization in the original DEF model has now been ruled out by pulsar-timing observations \cite{Damour:1996ke, Freire:2012mg, Shao:2017gwu, Kramer:2021jcw, Zhao:2022vig}, similar-type effects have been found in other contexts, including massive scalar fields \cite{Chen:2015zmx, Ramazanoglu:2016kul, Sperhake:2017itk, Sagunski:2017nzb}, more general scalar-tensor theories \cite{Andreou:2019ikc}, charged black holes~\cite{Herdeiro:2018wub, Fernandes:2019rez, Ikeda:2019okp}, higher-spin fields \cite{Ramazanoglu:2017xbl, Silva:2021jya, Ramazanoglu:2018hwk, Ramazanoglu:2018tig, Ramazanoglu:2019gbz, Barton:2021wfj}, as well as neutron stars \cite{Xu:2021kfh} and black holes in scalar-tensor-Gauss-Bonnet gravity, in which scalarization can be either curvature-induced \cite{Doneva:2017bvd, Doneva:2017duq, Silva:2017uqg, Antoniou:2017acq, Antoniou:2017hxj, Doneva:2018rou, Brihaye:2019kvj, Macedo:2019sem, Cunha:2019dwb,Collodel:2019kkx, East:2021bqk} or spin-induced \cite{Dima:2020yac, Herdeiro:2020wei, Berti:2020kgk, Hod:2020jjy, Doneva:2020nbb, Doneva:2021dqn}.

In the original setup of the DEF theory, scalarization was found to be potentialized in a dynamical setting.
In Ref.~\cite{Barausse:2012da}, it was shown that two neutron stars which were not compact enough to scalarize in isolation could scalarize dynamically in a close binary system.
A similar phenomenon, though different mechanisms are involved, has been observed in scalar-Gauss-Bonnet theories of gravity \cite{Silva:2020omi,Elley:2022ept,Annulli:2021lmn}.
In the case of DEF theory, dynamical scalarization (DS) was demonstrated in fully nonlinear numerical evolutions, but it appears already in numerically constructed quasi-equilibrium solutions \cite{Taniguchi:2014fqa,Shibata:2013pra}, and can be understood as a result of a feedback mechanism between the two neutron stars \cite{Palenzuela:2013hsa}, or more thoroughly, in the context of a resumed post-Newtonian expansion \cite{Sennett:2016rwa}.

A particularly elegant way to model DS is by an effective field theory for compact binaries~\cite{Goldberger:2004jt, Goldberger:2006bd, Goldberger:2009qd} that systematically takes into account the various scales in the problem, including nonperturbative effects in the strong-field regime close to the compact objects.
Compact bodies are modeled by point particles moving along a worldline in the effective theory.
Their internal dynamics, in particular from oscillation modes, is encoded in dynamical variables evolved along the worldlines.
This allows for the incorporation of scalar oscillation modes; scalarization arises when such modes become linearly unstable~\cite{Sennett:2017lcx, Khalil:2019wyy}.
DS is in fact modeled similarly to dynamical tides~\cite{Steinhoff:2016rfi,Gupta:2020lnv,Flanagan:2007ix}, where fluid oscillation modes are incorporated into the effective theory.
See also Refs.~\cite{Bernard:2019yfz, Creci:2021rkz} for further similarities between tidal effects and finite-size corrections in scalar-tensor theories.

While the effective action approach to DS was first formulated specifically for DEF theory~\cite{Sennett:2017lcx}, it can be straightforwardly generalized to other gravity theories containing scalar fields.
The main ingredient is a monopolar scalar mode close to the critical point of instability, which is included in the effective worldline action~\cite{Khalil:2019wyy}.
At leading order, the effective action contains only three parameters for each star.
These parameters must be matched to a specific gravity theory and compact object, but the form of the effective theory and hence its phenomenology is theory-independent.
Past work on this effective-action approach focused on an adiabatic (or quasi-stationary) analysis of the binary on quasi-circular orbits and its energetics, requiring only two of the three parameters of the action~\cite{Khalil:2019wyy}, finding good qualitative and quantitative agreement with numerical relativity~\cite{Sennett:2017lcx}.
In the present work, we study the full dynamical evolution of the binary and its scalar charges on generic (eccentric) orbits, where all three parameters become relevant.
This is an important step towards improving gravitational waveform models for dynamical-scalarization effects~\cite{Sampson:2013jpa, Sampson:2014qqa}.

This work is organized as follows.
In Sec.~\ref{sec:eft}, we recapitulate the effective action for DS.
Section~\ref{sec:binary_dynamics} develops the leading-order equations of motion for the binary and the scalar modes of the bodies, in particular the radiation-reaction (damping) forces.
The matching of all parameters of the effective action to neutron stars is performed in Sec.~\ref{sec:stt}, using the DEF class of scalar-tensor gravity as an example.
This allows for the investigation of the binary dynamics around the critical point of scalarization in Sec.~\ref{sec:results}.
Our conclusions are given in Sec.~\ref{sec:conclusions}, followed by a couple of Appendices providing detailed calculations on oscillation equations, as well as fluxes and radiation-reaction forces. We use units such that $c = G = 1$ throughout the text.

\section{Effective action model}
\label{sec:eft}

In this section, we briefly review the effective action model for DS of compact objects introduced in Refs.~\cite{Sennett:2017lcx,Khalil:2019wyy}.
We emphasize the fact that the following formalism is theory-independent~\cite{Khalil:2019wyy}.
Let us first recapitulate the widely used ``instantaneous'' effective action for a neutron star in scalar-tensor gravity given by a point particle moving along a worldline $y^\mu(\tau)$, which in the Einstein frame reads
\begin{equation}
S_\text{NS}^\text{inst} = - \int d\tau \, m_E (\varphi),
\end{equation}
where $m_E$ is the Einstein-frame mass depending on the external scalar field $\varphi(y^\mu)$.
This action is valid for an instantaneous response of the neutron star to the external scalar field $\varphi$, i.e., when internal relaxation timescales are much shorter than external timescales at which $\varphi$ varies.
Small corrections to this approximation can be incorporated through terms involving $\dot \varphi$.
However, when internal timescales exceed external ones, as for DS, additional dynamical variables representing the internal dynamics need to be evolved along the worldline. 

Hence, we include in the effective action a dynamical variable $q(\tau)$ representing the monopolar scalar oscillation mode that becomes linearly unstable at scalarization.
That is, the fundamental scalar mode has a frequency that vanishes (its period timescale diverges) at the critical point associated with the transition to a scalarized state.
For simplicity, we do not include further dynamical modes such as scalar-mode overtones or fluid oscillation modes in the effective theory (see, e.g., Refs.~\cite{Steinhoff:2016rfi,Gupta:2020lnv} for the latter), and neglect rotation of the star, but otherwise the model is rather generic or theory-independent.
Let us now consider a binary neutron star system. Around the critical point, the action describing the dynamics of $q$ can then be written as~\cite{Khalil:2019wyy}
\begin{equation} \label{eq:SCO}
S_\text{NS}^\text{crit} = \int d\tau \left[ \frac{c_{\dot{q}^2}}{2}\dot{q}^2 + \varphi(y) q - m(q) + O\left(\frac{R^2}{r^2}\right) \right],
\end{equation}
where $\dot~ = d / d \tau$ and the effective action is expanded in powers of the neutron-star size $R$ over the orbital separation $r$ of the binary.
Likewise, $m(q)$ can be expanded as
\begin{equation}\label{eq:mofq}
    m(q) = c_{(0)} - \varphi_0 q + \underbrace{\frac{c_{(2)}}{2!} q^2 + \frac{c_{(4)}}{4!} q^4 + O\left(\frac{R^2}{r^2}\right)}_{\displaystyle V(q)} ,
\end{equation}
where we include a possible cosmological value of the scalar field $\varphi_0$ (while $\varphi$ represents the field at the worldline which emanates from the binary companion);
the function $m(q)$ is even in $q$ if $\varphi_0 = 0$.
Note that $q^3 = O(R/r)$ close to the critical point where $c_{(2)}$ becomes small. 
The coefficient $c_{(0)}$ has the interpretation of the body's ADM mass when $q=0$ (i.e., in general relativity).
The Euler-Lagrange equation yields (suppressing the power counting in $R/r$ from now on)
\begin{equation} \label{eq:oscillator}
c_{\dot{q}^2} \ddot{q} + V'(q) = \varphi(y) + \varphi_0.
\end{equation}
A stationary-state solution ($\dot{q} \approx 0$) for $q$ can be employed if the external scalar field is not varying too rapidly.
In this case it holds $m(q) = m_E(\varphi) + \varphi q$, which can be understood as a Legendre transformation (note that $q = - dm_E/d\varphi$).

By computing the Hamiltonian at leading, Newtonian order, one obtains~\cite{Sennett:2017lcx,Khalil:2019wyy} 
\begin{equation}
\label{Hbinary}
H = m_A + m_B + \frac{p_{q,A}^2}{2 c_{\dot{q}^2,A}} + \frac{p_{q,B}^2}{2 c_{\dot{q}^2,B}}  + \frac{\bm{p}_A^2}{2m_A} + \frac{\bm{p}_B^2}{2 m_B} - \frac{m_A m_B}{r} - \frac{q_A q_B}{r},
\end{equation}
where $A$, $B$ label the two bodies in the binary, $\bm{p}_{A/B}$ are their linear momenta, $r$ is the interbody distance, and $p_{q,A/B}$ are the canonical conjugates to the variables $q_{A/B}$. 
To obtain an approximate solution for the scalar charge, we restrict to the (quasi-)stationary case $p_{q,A/B} \approx 0$, then the equation of motion for $q_A$ reads
\begin{equation}
0 = \frac{\partial H}{\partial q_A} = z_A \left(- \varphi_0 + c_{(2),A} q_A + \frac{c_{(4),A}}{6} q_A^3 \right) - \frac{q_B}{r},
\end{equation}
where $z_A \equiv \partial H/\partial m_A =  1 - \bm{p}_A^2/(2m_A) - m_B/r$ is the redshift of body $A$. 
Assuming, for simplicity, that $\varphi_0 = 0$, that the two bodies are identical ($q \equiv q_A = q_B$), and that we can neglect post-Newtonian corrections to the redshift ($z_A \approx 1$), one obtains 
\begin{equation}
0 = \frac{\partial H}{\partial q} = - 2 q \left(\frac{1}{r} - c_{(2)} - \frac{c_{(4)}}{6} q^2 \right).
\end{equation}
There are three solutions for the equation above, the trivial one, with $q= 0$, and 
\begin{equation}
\label{qAnalytic}
q = \pm \sqrt{\frac{6}{c_{(4)}} \left( \frac{1}{r} - c_{(2)}\right)}.
\end{equation} 
The stability condition, $\partial^2 H/\partial q^2 \geq 0$, is violated for the trivial solution if $1/r > c_{(2)}$. This is fulfilled for all $r$ if $c_{(2)} < 0$, which corresponds to the case of spontaneous scalarization. DS is captured in this setup by the fact that even when $c_{(2)} >0$ the trivial solution can become unstable for sufficiently small binary separations.
In the present work, we go beyond the quasi-stationary case and investigate the dynamical evolution of $q$ around the moment of DS.

A similar analysis could be performed for dipolar ($\ell =2$) or generic $\ell$-polar scalar oscillation modes, with an interaction potential $\sim r^{-2\ell-1}$ in the Hamiltonian and hence a scalar-polarization condition $r^{-2\ell-1} > \text{const}$.
However, since the monopolar $\ell=0$ modes typically have a lower frequency than the higher $\ell>0$ modes, it is expected that a monopolar mode scalarizes before this condition is reached for the higher modes.
Still, adding a dynamical dipolar mode to the model can be interesting, since it can be resonantly driven by the monopolar scalar charge of the companion, with associated contributions to the dipolar scalar radiation (analogous to dynamical tidal effects~\cite{Steinhoff:2016rfi,Flanagan:2007ix}).

Let us elaborate on the matching of the coefficients $c_{\dot{q}^2}$, $c_{(n)}$ of the effective theory to the properties of the compact object obtained by solving the inner problem.
If one restricts to equilibrium configurations, with $\dot{q} = 0$, the procedure described in Ref.~\cite{Khalil:2019wyy} reduces operationally to the one presented for scalar-tensor theories in Ref.~\cite{Sennett:2017lcx}. In particular, one can obtain the coefficients $c_{(2)}$ and $c_{(4)}$ through the following steps:
\begin{itemize}
\item[(i)] Solve (numerically) the relevant structure equations for an isolated object in the full theory, computing a sequence of solutions at a fixed baryon mass $M_b$ and for different values of the asymptotic scalar field $\varphi_\infty$, then extract the Einstein-frame ADM mass $m_E(\varphi_\infty)$ and scalar charge $q(\varphi_\infty)$. Note that for each value of $\varphi_\infty$ more than one equilibrium solution may exist.

\item[(ii)] Compute $m(q)$ through
\begin{equation} \label{eq:mofq2}
m(q) = m_E(\varphi_\infty) + \varphi_\infty q(\varphi_\infty),
\end{equation}
for each value of $\varphi_\infty$ and each possible equilibrium solution.

\item[(iii)] Fit the polynomial $V(q)$ in Eq.~\eqref{eq:mofq} to the numerical values of $m(q)-c_{(0)}$, with $\varphi_0 = 0$, and extract the quadratic $c_{(2)}$ and quartic $c_{(4)}$ coefficients, which encode the existence of spontaneous and dynamical scalarization, as described above.
\end{itemize}

We can additionally match the coefficient $c_{\dot{q}^2}$ by making the connection to the oscillation frequency of the fundamental scalar mode.
For this purpose, we specialize Eq.~(\ref{eq:oscillator}) to the force-free case $\varphi + \varphi_0 = 0$ and to small oscillations of frequency $\omega_0$ around the nonscalarized equilibrium $q=0$,
\begin{equation}
\ddot{q} = - \frac{c_{(2)}}{c_{\dot{q}^2}} q = - \omega_0^2 q .
\end{equation}
However, this picture is not complete, since $q$ also sources monopolar scalar radiation and should include a radiation-reaction force (a damping term involving $\dot{q}$) in the oscillator equation, which is derived in Appendix~\ref{app:cqd2}.
Now, the parameters in our damped harmonic oscillator $q$ can be matched to the (complex) quasi-normal mode frequency $\omega_\varphi$ obtained from neutron-star perturbation theory, resulting in $\omega_0^2 = |\omega_\varphi|^2 = \Re[\omega_\varphi]^2 + \Im[\omega_\varphi]^2$, which was worked out in detail within a different context in Ref.~\cite{Maggiore:2007nq}.
We can hence express the remaining coefficient as
\begin{equation}
\label{eq:cq_special}
c_{\dot{q}^2} = \frac{c_{(2)}}{|\omega_\varphi|^2}.
\end{equation}

A different match for $c_{\dot{q}^2}$ can be obtained by comparing damping times from radiation reaction (Appendix~\ref{app:cqd2}) and the quasi-normal mode frequency.
While this alternative matching is only approximate (with our radiation-reaction force based on a weak-field expansion), we find similar results for $c_{\dot{q}^2}$, at least for the specific model explored in Secs.~\ref{sec:stt} and \ref{sec:results}.

Finally, one can also expand Eq.~(\ref{eq:oscillator}) around a spontaneously scalarized solution $q_0$ such that $V'(q_0) = 0$, or $q_0 = \pm \sqrt{- 6 c_{(2)} / c_{(4)}}$.
Plugging $q = q_0 + \bar{q}$ in Eq.~\eqref{eq:oscillator} and $\varphi + \varphi_0 = 0$, and expanding to linear order in $\bar{q}$, yields
\begin{equation}
c_{\dot{q}^2} \ddot{\bar{q}} \simeq - \left(c_{(2)} + \frac{c_{(4)}}{2} q_0^2\right) \bar{q} ,
\end{equation}
from which we obtain the matching
\begin{align}
\label{eq:cq_general}
c_{\dot{q}^2} &= \frac{1}{|\omega_\varphi|^2} \left( c_{(2)} + \frac{c_{(4)}}{2} q_0^2\right) \nonumber\\
 &= - \frac{2 c_{(2)}}{|\omega_\varphi|^2} \qquad \text{(around scalarized solution $q_0$).}
\end{align}

\section{Binary dynamics}
\label{sec:binary_dynamics}

In this section, we obtain the set of equations governing the dynamics of a compact binary under the effective action model described in the previous section.
Notice that this formalism is still theory-independent. 

In the center-of-mass frame, and using polar coordinates, the Hamiltonian for a binary at leading order, Eq.~\eqref{Hbinary}, can be written as
\begin{equation}\label{Hamiltonian}
H = m_A + m_B
+ \frac{\bm{p}^2}{2 \mu} 
+ \frac{p_{q,A}^2}{2 c_{\dot{q}^2,A}} + \frac{p_{q,B}^2}{2 c_{\dot{q}^2,B}} 
- \frac{M\mu}{r} - \frac{q_A q_B}{r},
\end{equation}
where the center-of-mass momentum $\bm{p}=\bm{p}_A=-\bm{p}_B$, $\bm{p}^2=p_r^2 + L^2 / r^2$, where $L$ is the orbital angular momentum of the system, and the total and reduced masses are defined by 
\begin{equation}
M = m^0_A + m^0_B, \qquad 
\mu = \frac{m^0_A m^0_B}{M},
\end{equation}
i.e., in terms of the constant masses $m^0_{A/B} = c_{(0),A/B}$ to keep the equations of motion at leading order.
These equations are given by
\begin{alignat}{2}\label{EOMs}
\dot{r} &= \frac{\partial H}{\partial p_r}, \qquad
&\dot{p}_r &= - \frac{\partial H}{\partial r} + \mathcal{F}_r^\text{quad} + \mathcal{F}_r^\text{dip}, \nonumber\\
\dot{\phi} &= \frac{\partial H}{\partial L}, \qquad
&\dot{L} &= - \frac{\partial H}{\partial \phi} + \mathcal{F}_\varphi^\text{quad} + \mathcal{F}_\varphi^\text{dip}, \nonumber\\
\dot{q}_A &= \frac{\partial H}{\partial p_{q,A}}, \qquad
&\dot{p}_{q,A} &= - \frac{\partial H}{\partial q_A} + \mathcal{F}_{q_A}^\text{mon}, \nonumber\\
\dot{q}_B &= \frac{\partial H}{\partial p_{q,B}}, \qquad
&\dot{p}_{q,B} &= - \frac{\partial H}{\partial q_B} + \mathcal{F}_{q_B}^\text{mon}, 
\end{alignat}
where we add the leading-order tensor quadrupole, scalar dipole, and scalar monopole radiation-reaction forces;
the dipole force is expected to dominate in the scalarized phase for binaries with unequal masses and charges, while the quadrupole dominates in the unscalarized phase.

For the tensor quadrupole radiation-reaction force, we use the leading-order expressions given by Eqs. (3.67) and (3.68) in Ref.~\cite{Bini:2012ji}, which read
\begin{align}
\mathcal{F}_r^\text{quad} &= \frac{8}{15} \frac{M\mu^2}{r^3} \dot{r} \left(21 r^2 \dot{\phi}^2 - \frac{M}{r}\right), \nonumber\\
\mathcal{F}_\varphi^\text{quad} &= - \frac{8}{15} \frac{M\mu^2}{r} \dot{\phi} \left(-\dot{r}^2 + 2r^2\dot{\phi}^2 + 2\frac{M}{r}\right).
\end{align}
The monopole and dipole radiation-reaction forces are derived in Appendix~\ref{app:RR}, and are given by
\begin{align}
\mathcal{F}_{q_A}^\text{mon} &= \mathcal{F}_{q_B}^\text{mon} = - \dot{q}_A(t) - \dot{q}_B(t), \\
\mathcal{F}_r^\text{dip} &= \frac{2}{3} \frac{M}{\mu r^3} p_r \left(\frac{m_B^0}{M} q_A - \frac{m_A^0}{M} q_B\right)^2  \left(1 + \frac{q_A q_B}{M\mu}\right), \nonumber\\
\mathcal{F}_\varphi^\text{dip} &= - \frac{1}{3}\frac{ML}{\mu r^3} \left(\frac{m_B^0}{M} q_A - \frac{m_A^0}{M} q_B\right)^2 \left(1 + \frac{q_A q_B}{M\mu}\right).
\end{align}
Note that the monopole radiation-reaction force enters the equations of motion for the scalar charges, but not the orbital equations, since the monopole energy flux depends only on $\dot{q}$ but not on the orbital variables.

In order to set initial conditions, the following relations are useful:
\begin{equation}
  r = \frac{k}{1+e\cos\phi} , \qquad k = a (1 - e^2) = \frac{L^2}{\mu^2 M}\,,
\end{equation}
where $k$ is the semilatus rectum, $a$ the semi-major axis, and $e$ the eccentricity of the orbit.
The periastron distance is then $r_p = k / (1+e)$ and the apastron distance is $r_a = k / (1-e)$, with the eccentricity $e = (r_a - r_p) / (r_a + r_p)$.
If we start the orbit at apastron and assume that the binary starts unscalarized, then the initial conditions read
\begin{equation}
  \phi = \pi, \qquad r = r_a, \qquad L = \mu \sqrt{r_a M (1-e)}\,.
\end{equation}

In the absence of radiation reaction, $p_r=0$ at apastron. However, for slowly varying $L$ due to radiation, and for circular orbits, the initial condition for $p_r$ reads
\begin{equation}
  p_r = \mu \dot r = \frac{2 L}{\mu M} \dot{L} = \frac{2 L}{\mu M} \mathcal{F}_\varphi^\text{quad} = - \frac{64}{15} \frac{\mu^2 M^2}{r^3}.
\end{equation}
This also provides a good approximation for eccentric orbits at apastron.

For the scalar charge, in the presence of a (small) cosmological scalar field $\varphi_0$, and assuming stationarity, $\dot{q} = 0$, one gets from Eq.~\eqref{eq:oscillator} that
\begin{align}
c_{(2),A} q_A + O(q_A^3) &= \varphi_0 + \frac{q_B}{r} , \nonumber\\ c_{(2),B} q_B + O(q_B^3) &= \varphi_0 + \frac{q_A}{r},
\end{align}
leading to
\begin{equation}
\label{initq}
 q_A = \frac{\varphi_0}{c_{(2),A}} \frac{1 + \frac{1}{c_{(2),B} r}}{1 - \frac{1}{c_{(2),A}c_{(2),B}r^2}} \approx \frac{\varphi_0}{c_{(2),A}},
\end{equation}
and similarly for $q_B$.
Notice that the last approximation only holds away from the scalarization point, since $c_{(2),B} r \sim 1$ close to scalarization.

Given that we start the evolution in the unscalarized regime, the initial value of $p_q$ is zero. However, if we set the cosmological scalar field $\varphi_0=0$, one needs to take $p_q$ as a small but nonzero number to perturb the system away from the unstable solution, $q=0$, and allow the transition to the DS regime.

\section{A particular scalar-tensor model}
\label{sec:stt}

In this section, we illustrate how to compute the effective action coefficients $c_{(2)}$, $c_{(4)}$, and $c_{\dot{q}^2}$ for the case of a massless scalar-tensor theory defined by the (Einstein-frame) action
\begin{equation}
	S = \frac{1}{16\pi} \int{d^4 x \sqrt{-g} \left( \mathcal{R} 
	- 2 \nabla_\mu \varphi \nabla^\mu \varphi \right)}
	+ S_\text{m}[\Psi_\text{m} ; a(\varphi)^2 g_{\mu\nu} ],
\label{eq:general_action}
\end{equation}
where $g\equiv \det(g_{\mu\nu})$ and $\mathcal{R}$ is the Ricci scalar. 
The function $a(\varphi)$---which defines the Jordan-frame metric $\tilde{g}_{\mu\nu} \equiv a(\varphi)^2 g_{\mu\nu}$ to which matter fields $\Psi_\text{m}$ couple universally---is fixed to 
\begin{equation} \label{eq:aDEF}
    a(\varphi) = \exp(\beta \varphi^2/2).
\end{equation}
This model, introduced by Damour and Esposito-Far\`ese \cite{Damour:1993hw}, is arguably the simplest one displaying spontaneous scalarization, and most works on the subject revolve around it. Although most (or all) of the range $\beta \lesssim -4.5$ allowing for spontaneous scalarization in this model \cite{Harada:1997mr,Novak:1998rk,Silva:2014fca,AltahaMotahar:2017ijw} has now been ruled out by pulsar timing observations \cite{Damour:1996ke, Freire:2012mg, Shao:2017gwu, Kramer:2021jcw, Zhao:2022vig}, it is still a good prototype for our discussion. We have analyzed both the cases where $\beta = -5$ and $-6$, and found very similar behaviors. In what follows, results for $\beta = -5$ will be displayed. Additionally, the stellar fluid will be described by a two-piece polytrope with adiabatic index $\Gamma_1 = 3$ in the core, and $\Gamma_1 = 1.3$ in the crust, with the transition happening at $1.66 \times 10^{13}$g/cm$^3$; this is the same equation of state adopted for the computation of radial mode frequencies in Ref.~\cite{Mendes:2018qwo}, which will be used in what follows.

We note that there is an interesting range, i.e., $-4.5 \lesssim \beta \lesssim -3.5$, which is not ruled out by binary-pulsar observations but still would allow for DS before the two NS merge \cite{Palenzuela:2013hsa}. However, for a fixed value of $\beta$ in that range, there is no scalarization critical point that can be approached parametrically, and around which one can safely assume the validity of the effective action. Still, there is no technical issue that prevents applying our model to that case as well, and we expect a similar phenomenological behavior.

\subsection{Potential coefficients: $c_{(2)}$ and $c_{(4)}$}
\label{sec:pA}

In order to feed the effective action model with the potential parameters $c_{(2)}$ and $c_{(4)}$, one must consider the ``inner'' problem of an isolated neutron star with some fixed baryon mass $M_b$, subject to an external (varying) scalar field---as per item (i) in Sec.~\ref{sec:eft}. The structure equations in this case are given, e.g., by Eqs.~(31)$-$(34) of Ref.~\cite{Mendes:2016fby}. Next, following item (ii), one computes $m(q)$ in Eq.~(\ref{eq:mofq2}) from the ADM mass $m_E(\varphi_\infty)$ and scalar charge $q(\varphi_\infty)$---in Ref.~\cite{Mendes:2016fby} these are denoted by $M$ and $\omega$, and are given in Eqs.~(37) and (38), respectively.
For baryon masses close to the critical value for the onset of spontaneous scalarization ($M_{b,cr} = 1.3474 M_\odot$ for the scalar-tensor model and equation of state described above), the potential $V(q) = m(q) - c_{(0)}$ [with $c_{(0)} = m(0)$] is well approximated by the truncated expansion
\begin{equation}
\label{eq:potential}
V(q) = \frac{c_{(2)}}{2} q^2 + \frac{c_{(4)}}{4!} q^4.
\end{equation}
The final step (iii) consists in extracting the coefficients $c_{(2)}$ and $c_{(4)}$ of the best fit to the numerical data.

\begin{figure}[t]
    \includegraphics[width=\linewidth]{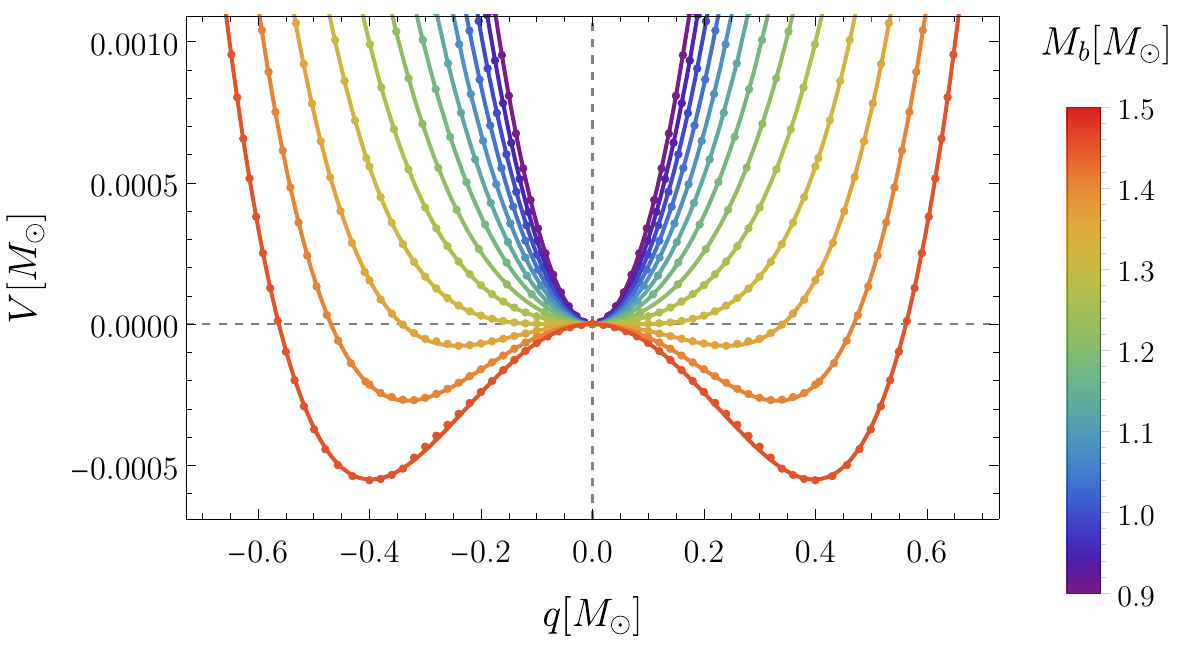}
    \caption{$V(q)$ for fixed baryon masses (ranging from $0.9$ to $1.5 M_\odot$). Points represent the numerically computed values, which are fitted by polynomial expressions of the form (\ref{eq:potential}). Fit coefficients are represented in Fig.~\ref{fig:c24}.}
    \label{fig:potential}
\vspace{\floatsep}
    \includegraphics[width=0.95\linewidth]{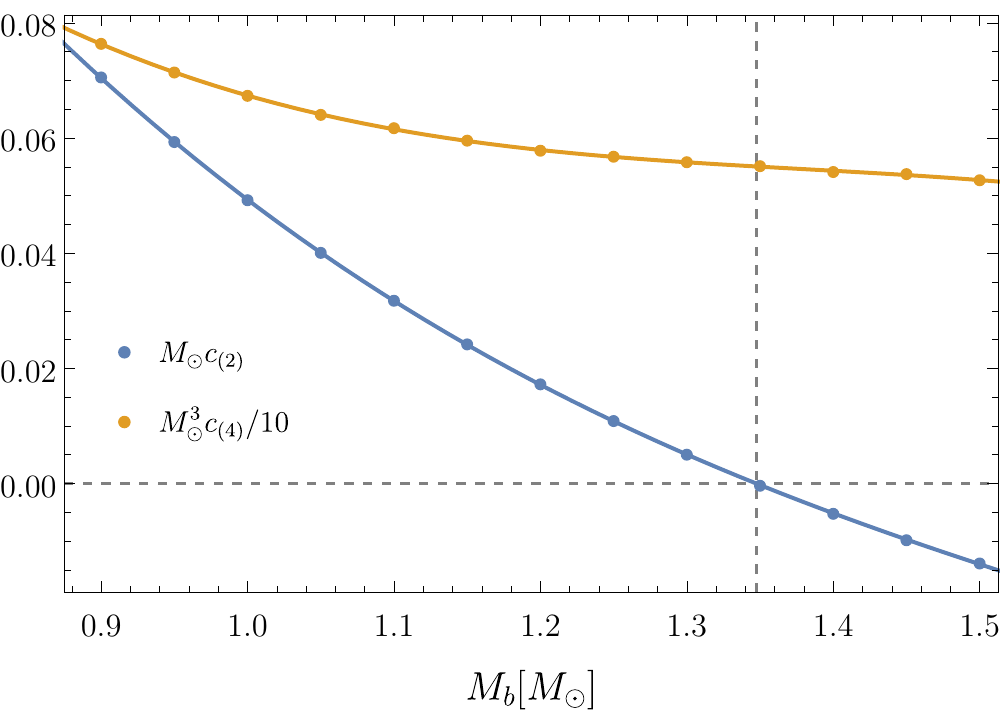}
    \caption{Coefficients $c_{(2)}$ and $c_{(4)}$ for some baryon masses around the critical point (represented by the dashed vertical line). Points represent the numerically computed values, which are fitted by expressions (\ref{eq:c2interp}) and (\ref{eq:c4interp}).}
    \label{fig:c24}
\end{figure}

Figure \ref{fig:potential} illustrates the potential $V(q)$ for some baryon masses around $M_{b,cr}$. Before the critical point, $c_{(2)}>0$, the potential has a single minimum at $q = 0$. After the critical point, $c_{(2)}<0$, the potential has a local maximum at $q=0$ (corresponding to a GR-like unstable equilibrium solution) and two local minima $q \neq 0$ with opposite signs (corresponding to stable scalarized solutions).
The local extrema of $V(q)$ correspond to $\varphi_\infty = 0$, since from Eq.~\eqref{eq:mofq2}, $V'(q)=\varphi_\infty$.

In Fig.~\ref{fig:potential}, the numerically computed points are fitted by a polynomial expression of the form (\ref{eq:potential}), with the coefficients $c_{(2)}$ and $c_{(4)}$ represented in Fig.~\ref{fig:c24}. Their dependence on the baryon mass is well captured by the following polynomial fits: 
\begin{align}\label{eq:c2interp}
M_\odot c_{(2)} &\approx (x - x_{cr}) \left(-0.3796 + 0.3294 x - 0.09174 x^2 \right), \\
M_\odot^3 c_{(4)} &\approx 4.270 - 7.804 x + 5.545 x^2 - 1.337 x^3,\label{eq:c4interp}
\end{align}
where $x \equiv M_b/M_\odot$ and we set $x_{cr} = 1.3474$.

\subsection{Scalar modes and the coefficient $c_{\dot{q}^2}$}
\label{sec:phi_modes}

\begin{figure}[t]
    \includegraphics[width=0.95 \linewidth]{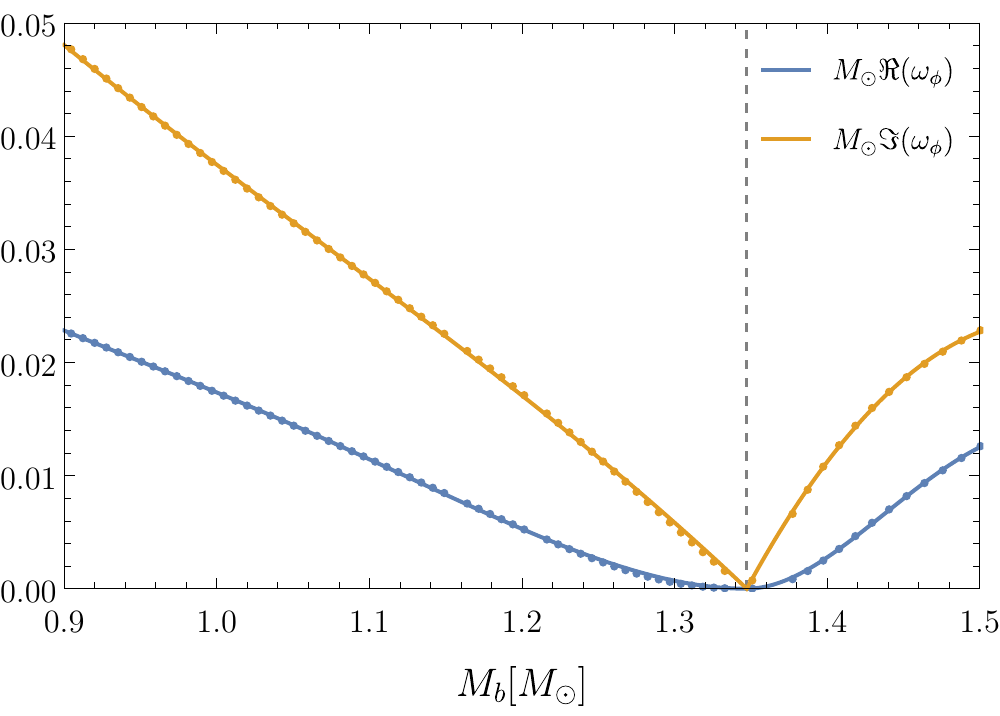}
    \caption{Real and imaginary parts of the fundamental scalar-led radial mode frequency (scaled by $M_\odot$). Points represent the numerically computed values, and lines represent the polynomial fits (\ref{eq:Romegainterp}) and (\ref{eq:Iomegainterp}). The dashed vertical line corresponds to the critical point, $M_{b,cr}$.}
\vspace{\floatsep}
    \label{fig:smodes}
    \includegraphics[width=0.95\linewidth]{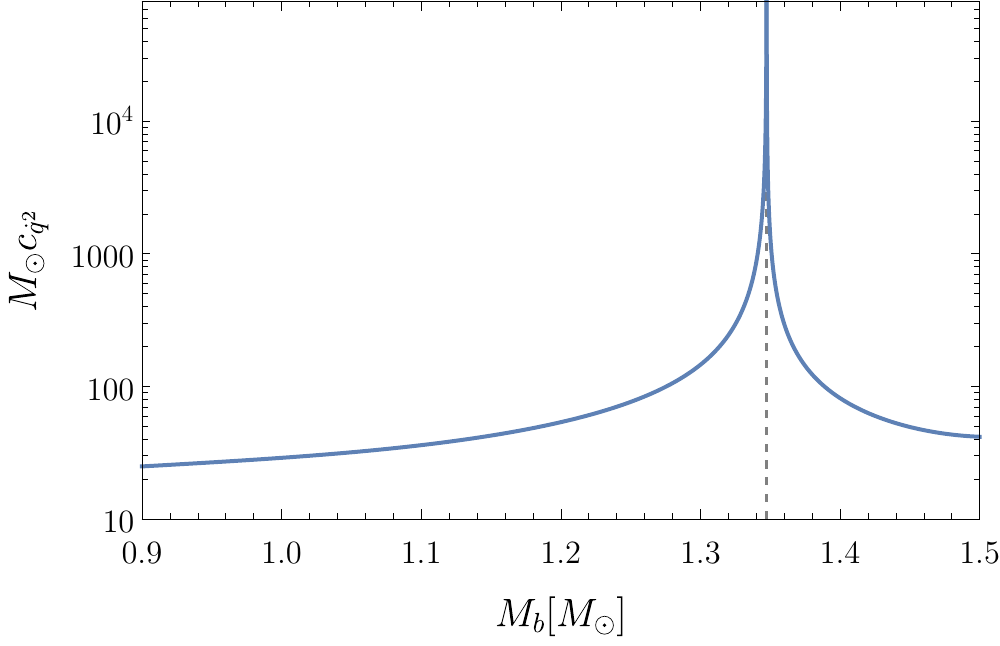}
    \caption{Coefficient $c_{\dot{q}^2}$ as a function of baryon mass around the critical point (displayed as a dashed vertical line).}
    \label{fig:cqd2}
\end{figure}

The coefficient $c_{\dot{q}^2}$ determines the strength of the kinetic term in Eq.~(\ref{eq:SCO}); in order to feed the model with this parameter, one can consider the dynamics of scalar field perturbations around a neutron star in equilibrium. As per Eqs.~(\ref{eq:cq_special}) and (\ref{eq:cq_general}), $c_{\dot{q}^2}$ depends on the potential coefficients $c_{(2)}$ and $c_{(4)}$, and on the frequency $\omega_0$, which encodes the dynamical timescale of scalar field oscillations. In this work, we adopt the following prescription: $\omega_0^2 \equiv \Re[\omega_\varphi]^2 + \Im[\omega_\varphi]^2$, where $\omega_\varphi$ is the fundamental scalar-led radial mode (or $\varphi$-mode) frequency.

The $\varphi$-mode frequency was computed in Ref.~\cite{Mendes:2018qwo} for stars subject to a vanishing asymptotic scalar field ($\varphi_\infty = 0$), and the result is reproduced in Fig.~\ref{fig:smodes} for the range of baryon masses considered previously ($0.9 M_\odot \leq M_b \leq 1.5 M_\odot$). For $M_b < M_{b,cr}$, i.e. before the onset of spontaneous scalarization, the (radial) perturbation equations for the scalar field and the fluid decouple, and the $\varphi$-modes are purely scalar perturbations. For $M_b > M_{b,cr}$, three equilibrium solutions exist. The trivial one, with $q=0$, is unstable under scalar field perturbations. Correspondingly, its $\varphi$-mode has $\Im(\omega) < 0$ and $\Re(\omega) = 0$ (not shown in the plot). The two nontrivial solutions are scalarized, with opposite scalar charges and identical oscillation frequencies (shown in Fig.~\ref{fig:smodes}). These correspond to coupled scalar and fluid oscillations \cite{Mendes:2018qwo,Sotani:2014tua}.

\begin{figure*}[t]
\includegraphics[width=\linewidth]{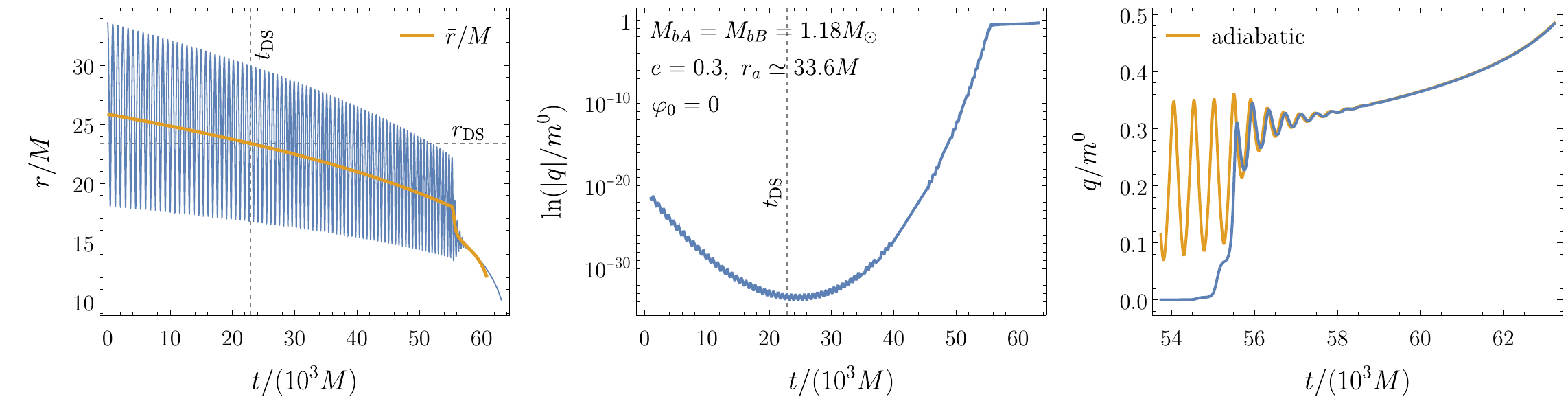}
\caption{Separation and scalar charge for an equal-mass binary with baryon mass $M_{bA} = M_{bB} = 1.18 M_\odot$, initial eccentricity $e = 0.3$, initial separation $r_a \simeq 33.6 M$, and asymptotic scalar field $\varphi_0=0$. The right panel shows the scalar charge around the time it reaches saturation, and compares it with the approximate solution in Eq.~\eqref{qAnalytic}.}
\label{fig:ecc_ex}
\end{figure*}

Around the critical point, we obtain the following polynomial interpolations for $\Re(\omega_\varphi)$ and $\Im(\omega_\varphi)$ as a function of $x = M_b/M_\odot$:
\begin{equation} \label{eq:Romegainterp}
\begin{aligned}
&\Re(\omega_\varphi)(x) \approx  \\
&\left\{
\begin{array}{ll}
(x - x_{cr})^2 \,\left(0.4465 - 0.9698 x + 0.6670 x^2\right), \quad  x \leq x_{cr} \\
(x - x_{cr})^2 \,\left(18.753 - 20.205 x + 5.374 x^2\right), \quad  x \geq x_{cr}
\end{array}
\right.
\end{aligned}
\end{equation}
and
\begin{equation} \label{eq:Iomegainterp}
\begin{aligned}
&\Im(\omega_\varphi)(x) \approx  \\
&\left\{
\begin{array}{ll}
(x - x_{cr}) \,\left(-0.20769 + 0.21629 x - 0.11647 x^2\right), \, x \leq x_{cr} \\
(x - x_{cr}) \,\left(1.5439 - 1.2407 x + 0.20706 x^2\right), \quad  x \geq x_{cr}
\end{array}
\right.
\end{aligned}
\end{equation}

From Eq.~(\ref{eq:cq_special}), $c_{\dot{q}^2} = c_{(2)} / |\omega_\varphi|^2$ before the critical point, and from Eq.~(\ref{eq:cq_general}), $c_{\dot{q}^2} = - 2 c_{(2)}/|\omega_\varphi|^2$ after it.\footnote{A caveat here is that in the spontaneously-scalarized regime, one has to take into account the coupling between scalar and fluid oscillations, the latter requiring a more sophisticated effective action with further dynamical variables.}
Figure \ref{fig:cqd2} shows the coefficient $c_{\dot{q}^2}$ as a function of baryon mass, where the polynomial interpolations in Eqs.~(\ref{eq:c2interp}), (\ref{eq:Romegainterp}), and (\ref{eq:Iomegainterp}) were used. From these expressions and our prescription for $c_{\dot{q}^2}$, it is apparent that the coefficient diverges at the critical point. Consequences of this divergence are discussed in Sec.~\ref{sec:results_parameter_variation}. 

\section{Results for the binary dynamics}
\label{sec:results}

In this section, we show results for the binary dynamics in the context of scalar-tensor gravity of the DEF class with $\beta = -5$.
We start by considering two representative examples of an equal-mass binary on an eccentric orbit. 
Then, we show how each of the model parameters affects the scalar charge evolution.
After that, we study binaries in a quasi-circular-orbit inspiral; we compare the quasi-stationary approximation with the full dynamical evolution, and compare the binding energy and waveform with general relativity.
We finally consider the case of unequal masses to show the effect of the dipole flux.

\subsection{Eccentric orbits}\label{sec:results_eccentric_orbits}
\begin{figure*}[t]
\includegraphics[width=0.9\linewidth]{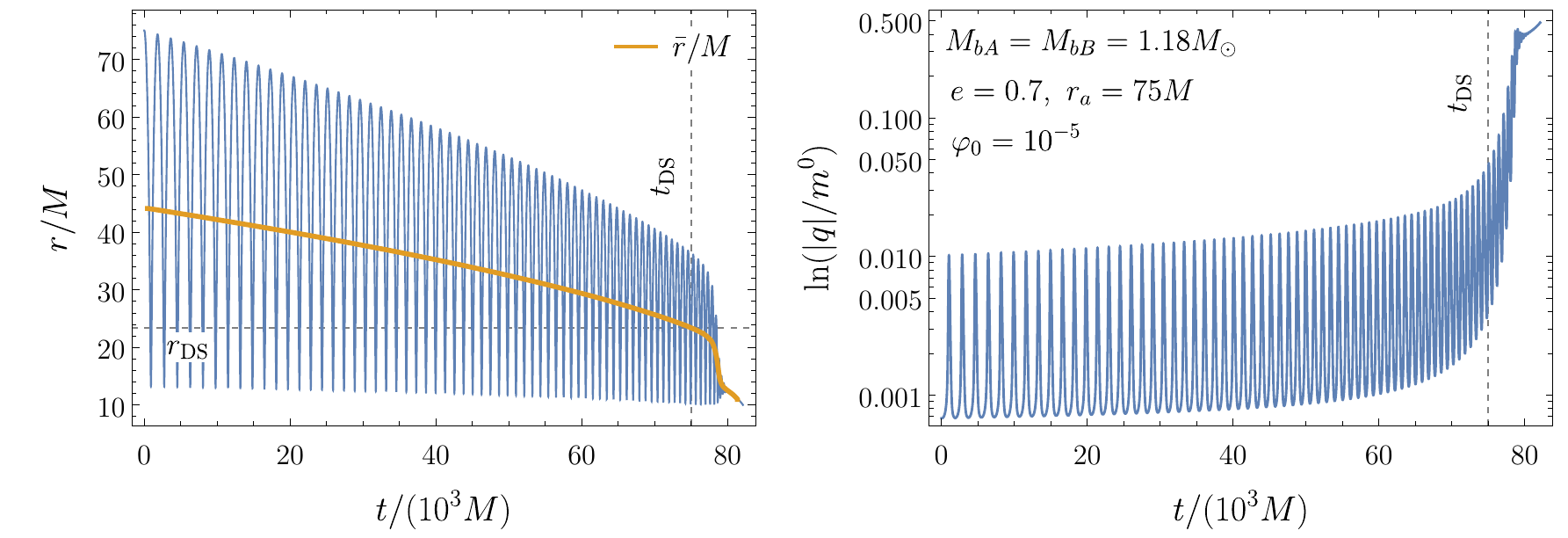}
\caption{Similar to Fig.~\ref{fig:ecc_ex} but for initial eccentricity $e = 0.7$ and asymptotic scalar field $\varphi_0=10^{-5}$. The local maxima of the scalar charge occur at periastron, which is below the expected separation allowing for DS.}
\label{fig:ecc_ex2}
\end{figure*}

We recall that, if we assume that the scalar charge evolves adiabatically, i.e., $p_q \approx 0$, then DS occurs for $r < r_\text{DS} \equiv 1/c_{(2)}$, where we denote the maximum radial separation allowing for DS by $r_\text{DS}$.
We also consider the mean radial separation $\bar{r}$, or the length of the semi-major axis, which we calculate by obtaining interpolating functions for the local maxima $r_a(t)$ and minima $r_p(t)$ of the solution $r(t)$, then evaluating
\begin{equation}
\bar{r}(t) \equiv \frac{r_a(t) + r_p(t)}{2}.
\end{equation}

Figure~\ref{fig:ecc_ex} shows an example for an equal-mass binary with baryon mass $M_{bA} = M_{bB} = 1.18 M_\odot$, initial eccentricity $e = 0.3$, and initial radial separation at apastron $r_a = r_\text{DS} + 22 M_\odot \simeq 33.6M$.
(Equal mass implies that the scalar charges $q_A$ and $q_B$ are identical, thus we refer to only one scalar charge $q$.)
We assume the asymptotic scalar field $\varphi_0 = 0$, which means the initial scalar charge, calculated from Eq.~\eqref{initq}, is zero. So we take $p_{q,A}= p_{q,B} = 10^{-20}$, representing some small perturbation to move the solution away from the unstable $q=0$ solution.
For all configurations considered in this paper, we stop the numerical evolution of the equations of motion at $r = 10 M$.

In the left panel of Fig.~\ref{fig:ecc_ex}, we plot $r(t)$ and $\bar{r}(t)$. The dashed horizontal line represents the radius $r_\text{DS}$ at which DS is expected, while the vertical line indicates the time $t_\text{DS}$ at which $\bar{r}$ crosses that radius.
The initial separation is such that $\bar{r}>r_{DS}$. Correspondingly, the scalar charge is initially dominated by an exponential damping, as shown in the middle panel.
Even though the periastron distance at the beginning is smaller than $r_\text{DS}$, the binary does not spend enough time inside $r_\text{DS}$ for DS to occur. 
However, the small oscillations in the scalar charge in this phase are due to the binary entering and exiting the DS region.

Shortly after $\bar{r}$ becomes smaller than $r_\text{DS}$, an exponential growth kicks in, and the scalar charge grows exponentially with small modulations.
It eventually saturates, and the saturation point of the exponential growth agrees with the quasi-stationary solution $q^2 \simeq 6/c_{(4)} [1/r(t) - c_{(2)}]$ from Eq.~\eqref{qAnalytic}, as can be seen in the right panel of Fig.~\ref{fig:ecc_ex}.
After reaching saturation, the increase in the scalar charge is proportional to $1/\sqrt{r}$.

In Fig.~\ref{fig:ecc_ex2}, we consider another example for a binary with the same baryon mass as in Fig.~\ref{fig:ecc_ex}, but with eccentricity $e=0.7$ and asymptotic scalar field\footnote{Note that in principle the scalar mode frequencies, and thus the value of $c_{\dot{q}^2}$, depend on $\varphi_0$, but were computed for $\varphi_0 = 0$. However, as with other macroscopic properties of the neutron star, we do not expect these mode frequencies to change appreciably in the range of (small) values of $\varphi_0$ considered in this section.} $\varphi_0 = 10^{-5}$.
We see that the high eccentricity and nonzero $\varphi_0$ cause the scalar charge to increase at periastron and decrease at apastron.
This behavior was first demonstrated in Ref.~\cite{Palenzuela:2013hsa} for eccentric binaries and was dubbed ``transient DS''.
The magnitude of the oscillations of the scalar charge increases with increasing eccentricity and for baryon masses closer to the critical value $M_{b,cr}$.
However, the increase in the scalar charge compared to its initial value, given by Eq.~\eqref{initq}, is mostly independent of $\varphi_0$, e.g., for the mass and eccentricity used in Fig.~\ref{fig:ecc_ex2}, the charge increases by about an order of magnitude regardless of the value of $\varphi_0$.

\subsection{Effect of the parameters on the scalar charge} \label{sec:results_parameter_variation}

\begin{figure*}[t]
\includegraphics[width=\linewidth]{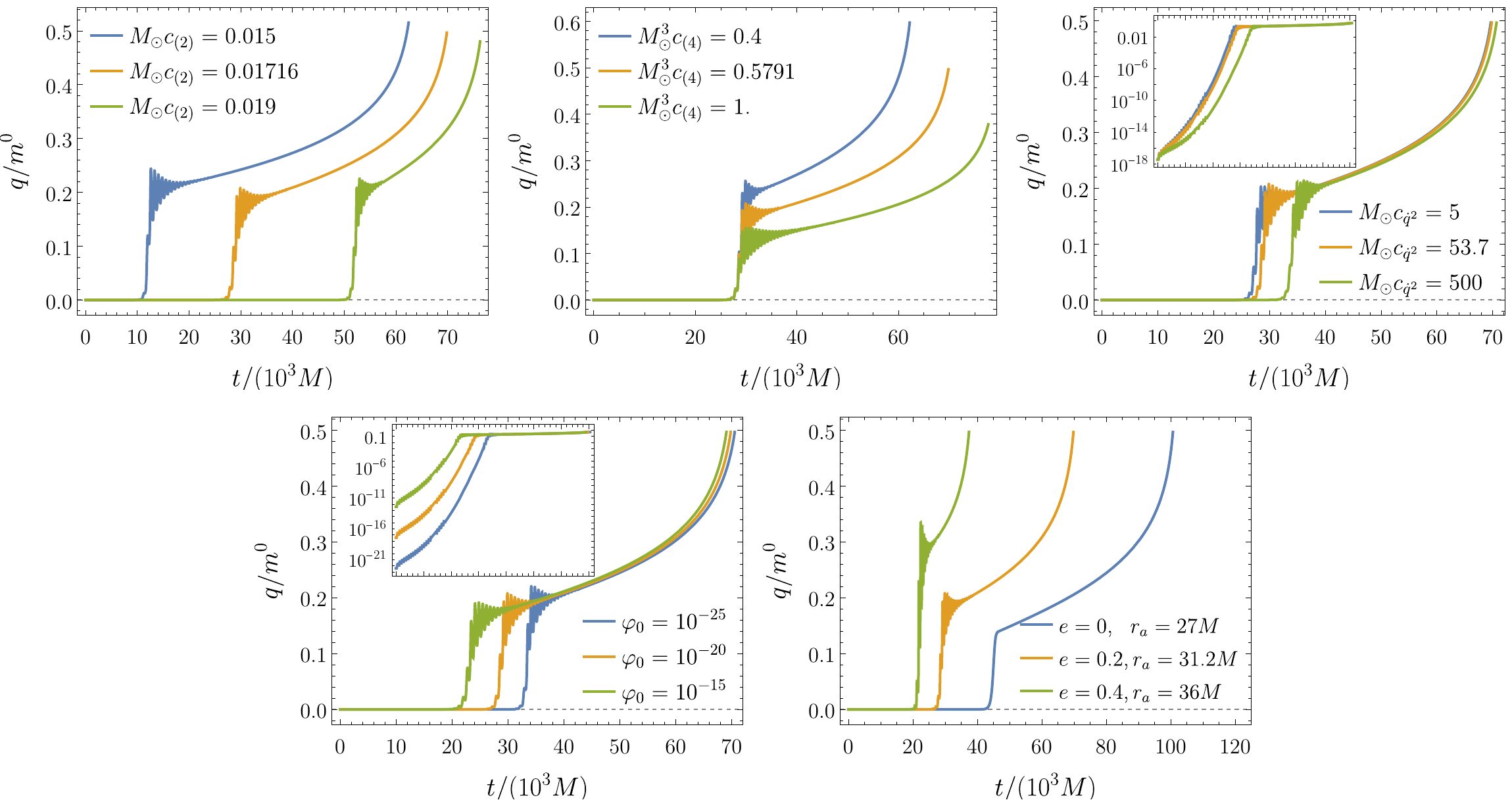}
\caption{Each panel shows the effect on the scalar charge of varying one of the parameters of the effective action ($c_{(2)}$, $c_{(4)}$, and $c_{\dot{q}^2}$), or varying the asymptotic scalar field $\varphi_0$ and eccentricity $e$.
The middle orange curve is the same in all panels, representing a configuration with baryon mass $M_{bA}=M_{bB}=1.2M_\odot$, initial eccentricity $e = 0.2$, initial separation $r_a \simeq 31.2 M$, and asymptotic scalar field $\varphi_0 = 10^{-20}$.}
\label{fig:params}
\end{figure*}

In order to explore the phenomenology of our theory-independent effective action, it makes sense to slightly move away from the specific effective parameters predicted by the DEF theory. In what follows, we focus on an equal-mass binary with baryon mass $M_{bA} = M_{bB} = 1.2M_\odot$, which would be characterized by the parameters $c_{(2)} = 0.01716$,  $c_{(4)} = 0.5791$, and  $c_{\dot{q}^2} = 53.7$ for the DEF model with $\beta = -5$ (cf. Sec.~\ref{sec:stt}).

To explore the effect of each parameter on the scalar charge, we vary one at a time and plot the charge in Fig.~\ref{fig:params}. For all panels in that figure, the middle curve corresponds the DEF values of those parameters. In the first three panels, we take the initial eccentricity as $e = 0.2$ and the asymptotic scalar field as $\varphi_0 = 10^{-20}$, and explore the effects of varying $e$ and $\varphi_0$ in the last two panels.
We start the evolution with initial separation such that $\bar{r}$ is just above the DS radius, so that the plots focus on the DS transition.
In particular, all curves in Fig.~\ref{fig:params} start at initial separation $r_a \simeq 31.2 M$, except for the last panel, in which we use $r_a = 36M$ for eccentricity $e = 0.4$ and $r_a= 27M$ for $e = 0$ for a better visualization.

From the figure, we observe the following:
\begin{itemize}
\item The coefficient $c_{(2)}$ has a significant effect on when scalarization occurs, since $r_\text{DS} = 1/c_{(2)}$, but it has a small effect on the magnitude of the charge after scalarization, as can be understood from Eq.~\eqref{qAnalytic}.

\item The coefficient $c_{(4)}$ affects the magnitude of the charge but not the DS radius, since from Eq.~\eqref{qAnalytic}, we see that the charge after scalarization is proportional to $1/\sqrt{c_{(4)}}$.

\item The coefficient $c_{\dot{q}^2}$ affects the time scale for the exponential growth of the scalar charge during scalarization, with smaller values leading to a shorter time.
Sufficiently increasing the value of $c_{\dot{q}^2}$, while keeping the other coefficients constant, would increase the scalarization timescale and prevent the binary from scalarizing before merger.

\item The asymptotic scalar field $\varphi_0$ changes the initial value of the scalar charge, and hence how early it reaches saturation. This is because $q \approx \varphi_0/c_{(2)}$ before scalarization.

\item The eccentricity $e$ affects the oscillations near saturation, with larger eccentricity leading to larger oscillations, because the binary spends more orbits going in and out of the DS region. 
\end{itemize}

Note that, as one approaches the critical mass for spontaneous scalarization, $c_{(2)}$ goes to zero, and $r_\text{DS}$ becomes arbitrarily large. On the other hand, the coefficient $c_{\dot{q}^2}$, which governs the timescale for  variations of the scalar charge, diverges at the critical point (cf.~Fig.~\ref{fig:cqd2}). These two effects compete in the initial phase of exponential growth of the scalar charge: as the critical point is approached, the growth of the scalar charge starts at earlier times due to the increase in $r_\text{DS}$ (favoring DS), but with a larger timescale due to the increase in $c_{\dot{q}^2}$ (disfavoring DS). Since the growth of the scalar field is exponential and the inspiral timescale is proportional to $r^4$ one expects DS to be favored as one moves closer to the critical point, which we have checked numerically. 

\subsection{Quasi-circular orbits}

\begin{figure}[h]
\includegraphics[width=\linewidth]{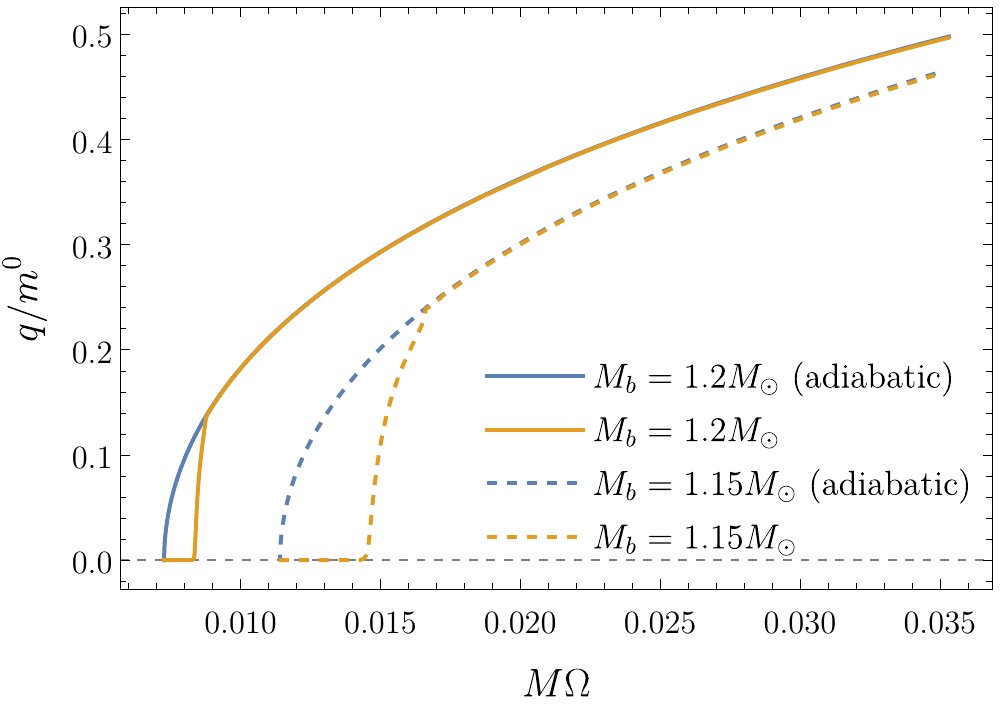}
\caption{Comparison of the quasi-stationary (adiabatic) approximation with the full dynamical evolution of the scalar charges, plotted versus the orbital frequency for equal-mass binaries on quasi-circular orbits.}
\label{fig:qcirc}
\end{figure}

\begin{figure*}[t]
\includegraphics[width=0.9\linewidth]{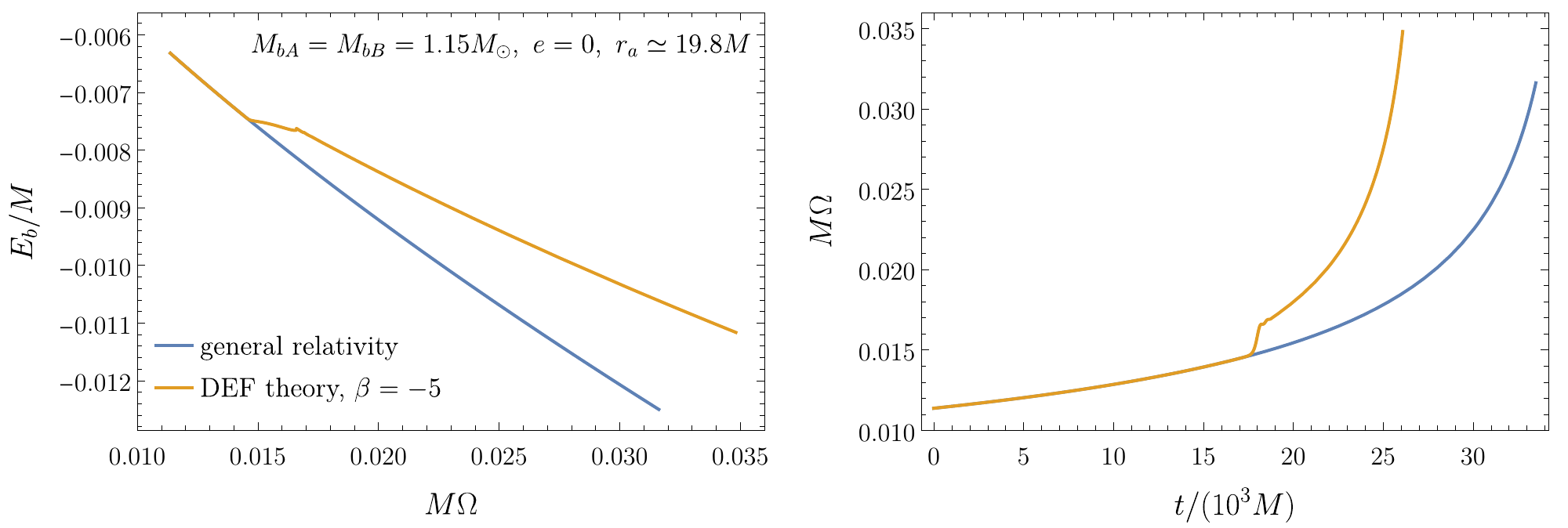}
\caption{Binding energy as a function of the orbital frequency (left panel) and orbital frequency as a function of time (right panel) for masses $M_{bA}=M_{bB}=1.15M_\odot$ in a quasi-circular inspiral.}
\label{fig:EbOmega}
\vspace{\floatsep}
\includegraphics[width=\linewidth]{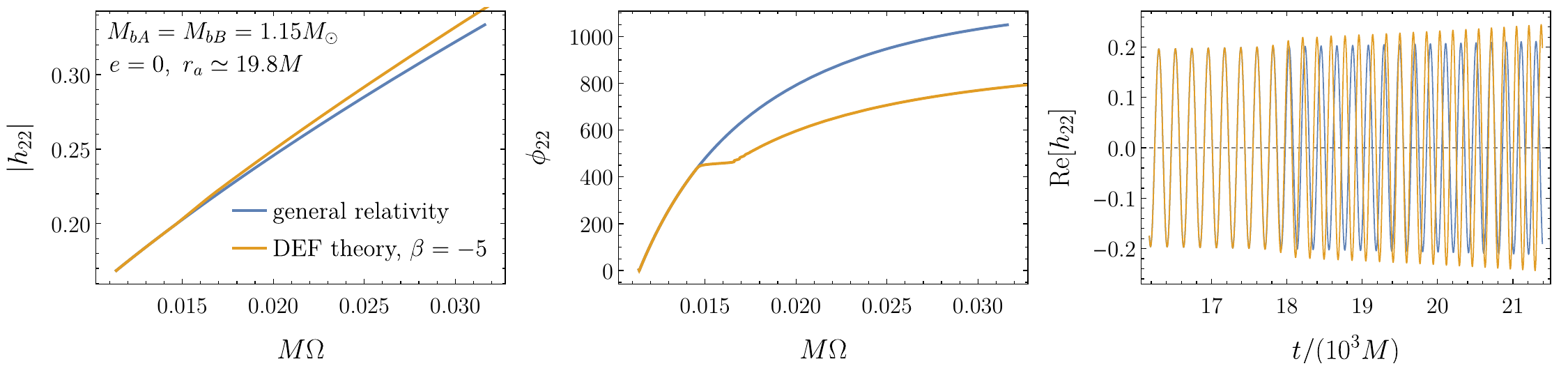}
\caption{The leading order of the (2,2) waveform mode for a quasi-circular inspiral.
The left and middle panels show the magnitude and phase of the waveform versus the orbital frequency, while the right panel shows the real part of the waveform around scalarization.}
\label{fig:waveform}
\end{figure*}

If the scalar charge is assumed to  change adiabatically ($\dot{q}\approx 0$), then the charge after scalarization is given by Eq.~\eqref{qAnalytic}.
In Fig.~\ref{fig:qcirc}, we compare that approximation with the numerical solution of Eqs.~(\ref{EOMs}), which include $\dot{q}$, for quasi-circular inspirals.

We consider two configurations: one with baryon mass $M_{bA} = M_{bB} = 1.2 M_\odot$, and the other with mass $M_{bA} = M_{bB} = 1.15 M_\odot$. For both, we use $\varphi_0 = 10^{-20}$, and start with initial separation $r_a = 1/c_{(2)} + 0.1M_\odot\simeq 19.8M$, which is just above the expected DS radius.
We see very good agreement between the analytical approximation and numerical solution, except at the beginning of the evolution, during the onset of DS. That is because including the $c_{\dot{q}^2}$ coefficient accounts for the time scale of the exponential growth of the scalar charge, and hence delays reaching the saturation point.

For the configuration with mass $M_{bA} = M_{bB} = 1.15 M_\odot$, we plot in Fig.~\ref{fig:EbOmega} the binding energy for the binary, which we define as the value of the Hamiltonian minus the constant ADM mass, that is
\begin{equation}
E_b \equiv H - M.
\end{equation}
To obtain $E_b$ for circular orbits as a function of the orbital frequency, we solve $\partial H/\partial r = 0$ and $\partial H/\partial L = \Omega$ for $r(\Omega)$ and $L(\Omega)$, with $p_r = 0$.
Substituting that solution into the Hamiltonian yields
\begin{align}
E_b &= m_A + m_B - M
+ \frac{p_{q,A}^2}{2 c_{\dot{q}^2,A}} + \frac{p_{q,B}^2}{2 c_{\dot{q}^2,B}} \nonumber\\
&\quad
-\frac{\mu}{2} (M \Omega )^{2/3} \left(1+\frac{q_A q_B}{\mu  M}\right)^{2/3}.
\end{align}
In Fig.~\ref{fig:EbOmega}, we see that the binding energy after scalarization decreases less rapidly than in general relativity, due to the sudden increase in the orbital frequency, causing the binary to become more tightly bound than in general relativity in a very short time.
This behavior qualitatively agrees with the quasi-equilibrium numerical calculations of Ref.~\cite{Taniguchi:2014fqa}.

We also plot in Fig.~\ref{fig:waveform} the Newtonian-order waveform, and compare it with general relativity.
The leading order of the $(2,2)$ waveform mode, for general orbits, is given by
\begin{equation}
h_{22} = - 4 \mu \sqrt{\frac{\pi}{5}} e^{-2i\phi} 
\left(\frac{M}{r} + \frac{L^2}{\mu^2 r^2} - \frac{p_r^2}{\mu^2} + 2 i \frac{L p_r}{\mu^2 r} \right),
\end{equation}
which can be written as a magnitude and phase $h_{22} \equiv |h_{22}| e^{i\phi_{22}}$.
We see from Fig.~\ref{fig:waveform} that the waveform agrees with general relativity until the onset of DS, which leads to a sudden change in the phase of the waveform, causing it to approach merger more quickly.

\begin{figure*}[t]
\includegraphics[width=0.9\linewidth]{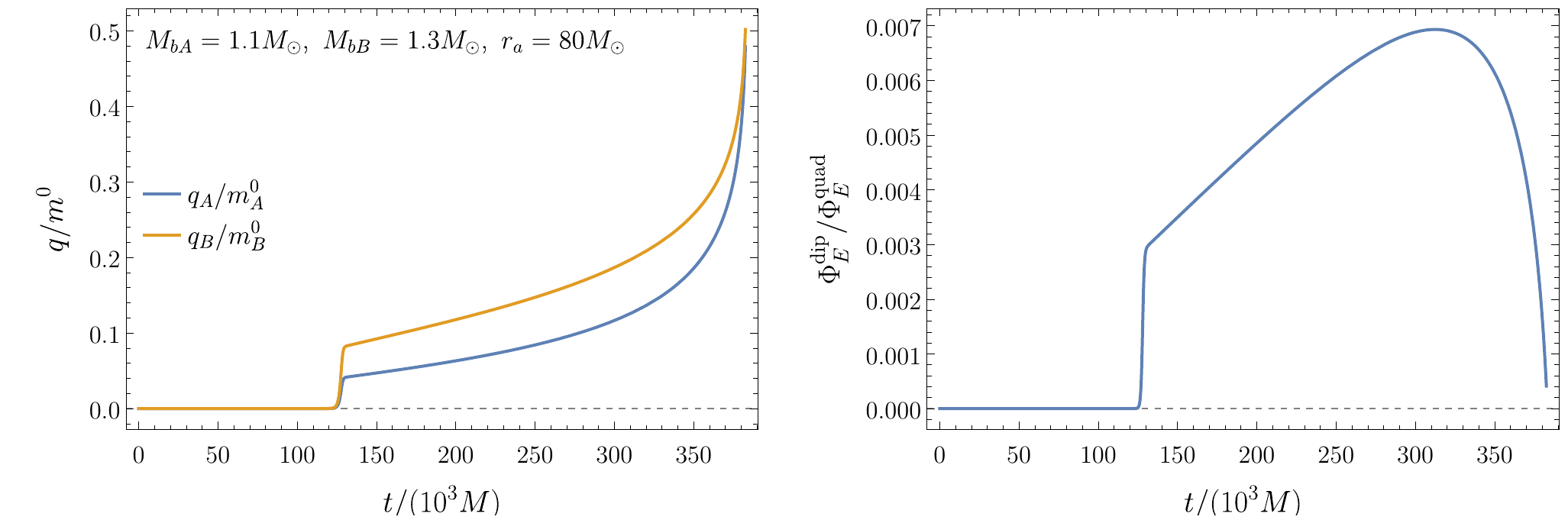}
\caption{Scalar charge (left panel) and the ratio of the scalar dipole flux to the tensor quadrupole flux (right panel), for a configuration with unequal masses on a quasi-circular orbit.}
\label{fig:dipFlux}
\end{figure*}

In addition to the tensor modes $h_{\ell m}$, scalar-tensor theories allow for scalar waveform modes $\psi_{\ell m}$, which were derived in Ref.~\cite{Bernard:2022noq} to 1.5PN order.
The dominant mode is $\psi_{11}$, and it is proportional to $(q_1/m_1 - q_2/m_2)$ at leading PN order. Therefore, the magnitude of $\psi_{11}$ is qualitatively similar to the leading order dipole flux, which is discussed in the following subsection.
The leading order of the $\psi_{00}$ mode is proportional to $q_1 + q_2$, so its behavior around scalarization is represented by the several plots in this paper for the scalar charge.

\subsection{Dipole flux}

So far, we have considered equal-mass systems, for which the leading-order dipole radiation vanishes [cf. Eq.~\eqref{dipFlux}].
To see the effect of the dipole flux, we consider a configuration with baryon masses $M_{bA} = 1.1  M_\odot$ and $M_{bB} = 1.3  M_\odot$, on a quasi-circular orbit with initial separation $r_a =80 M_\odot$, and asymptotic scalar field $\varphi_0 = 0$.

In Fig.~\ref{fig:dipFlux}, we plot the two scalar charges and see that they scalarize at almost the same time, with larger magnitude for the larger mass.
The right panel of that figure shows the ratio of the leading-order scalar dipole energy flux $\Phi_E^\text{dip}$, Eq.~\eqref{dipFlux}, to the tensor quadrupole flux, which is given for quasi-circular orbits by $\Phi_E^\text{quad} = 32 M^3\mu^2 / (5 r^5)$.
That ratio is of order $10^{-3}$ after scalarization, but decreases at small separations since $\Phi_E^\text{dip}$ is proportional to $1/r^4$ while $\Phi_E^\text{quad}$ is proportional to $1/r^5$.

\section{Conclusions}
\label{sec:conclusions}

Dynamical scalarization is a non-perturbative phenomenon that shows up in scalar-tensor theories of gravity, and involves the development of nontrivial scalar charges on inspiraling compact objects. Although this effect was originally discovered in fully nonlinear numerical simulations of neutron stars in scalar-tensor gravity~\cite{Barausse:2012da}, it is captured by compact-binary models that employ effective-field-theory methods~\cite{Sennett:2017lcx, Khalil:2019wyy}. By construction, this effective approach is designed to capture the UV and IR scales of the system, where the UV physics is simplified by modeling the binary components as point particles, while relevant internal degrees of freedom are captured by variables that evolve along the worldlines. In this way, the model can incorporate physical aspects such as tidal deformations and/or stellar oscillations.

Earlier effective-action models for DS considered the case where the evolution is quasi-stationary, and were restricted to quasi-circular orbits. In this work, we considered the full dynamical evolution predicted by the model for general (eccentric) orbits. This is achieved at leading order in the effective action, which contains only three parameters for each binary component. One of these parameters---the coefficient $c_{\dot{q}^2}$ of the kinetic term in Eq.~(\ref{eq:SCO})---was neglected in previous works dealing with the quasi-stationary case \cite{Khalil:2019wyy}. Here we suggested a prescription to match this coefficient to properties of the fundamental scalar mode of a neutron star, and considered its effect on the binary evolution. Additionally, radiation-reaction effects were incorporated at the level of the equations of motion. 
In order to show how the model works in practice, we chose to work with the well-known DEF scalar-tensor theory [cf.~Eqs.~(\ref{eq:general_action}) and (\ref{eq:aDEF})] with the parameter $\beta=-5$. We have thus worked out the full dynamics of eccentric, oscillatory binary neutron stars in DEF theory, near the phase transition to the scalarized regime.

The main effect of including the $c_{\dot{q}^2}$ term in the model is to account for the initial phase of exponential growth of the scalar charge, before it reaches saturation and the evolution becomes well described by the adiabatic solution. The time spent in the initial phase also depends on the cosmological value of the scalar field ($\varphi_0$), which determines the magnitude of the scalar charges prior to scalarization. $\varphi_0$ impacts the theory's parameterized post-Newtonian parameters~\cite{Will:2014kxa} and is thus subject to an upper bound from solar system observations; on the other hand, a lower bound is provided by the scalar field vacuum fluctuations.

General eccentric orbits were also considered. For such orbits, the scalar charge is amplified and suppressed successively as the binary enters and exits the DS radius---defined as the maximum radial separation allowing for DS in the stationary limit---, with the mean radial separation determining the overall behavior (see Fig.~\ref{fig:ecc_ex}).

By incorporating the effects of mode dynamics and dissipation, our work makes important steps towards improving the accuracy of post-Newtonian waveform models for compact binaries in gravity theories that allow for scalarization, and can thus be useful in constraining scalarization with gravitational-wave observations~\cite{Sampson:2013jpa, Sampson:2014qqa, Shao:2017gwu, Zhao:2019suc, Niu:2021nic, Guo:2021leu, Wong:2022wni}.

It is worth emphasizing that, although many of our results were obtained for a specific gravity theory, the effective action model introduced here is theory independent. Thus, it can be used as the basis for both theory-specific and theory-agnostic tests of DS. In the latter case, one could attempt to constrain directly the leading coefficients $(c_{(2)}, c_{(4)}, c_{\dot{q}^2})$ of the effective action---or combinations thereof, taking into account possible degeneracies in their relation to observables. Degeneracies related to uncertainties in the nuclear equation of state may also be relevant for the case of binary neutron star systems, but we note that scalarization also shows up for black holes in some scalar extensions of GR. In any case, waveform models for DS would benefit from higher-order post-Newtonian calculations in specific frameworks (see e.g.~Refs.~\cite{Mirshekari:2013vb,Bernard:2018hta, Bernard:2022noq} for efforts in a class of scalar-tensor models) and of possible refinements to the effective action, such as accounting for the interplay between fluid and scalar oscillation modes.

\section*{Acknowledgments}

We thank Robert Benkel for collaboration at an early stage of this work and for insightful discussions.
We also thank Hector O. Silva and the anonymous referee for useful comments on the manuscript.
N.O. acknowledges financial support by the CONACyT grants ``Ciencia de Frontera" 140630 and 376127, and by the UNAM-PAPIIT grant IA100721.
R.M. acknowledges partial funding from the National Council for Scientific and Technological Development (CNPq) and by the Carlos Chagas Filho Research Support Foundation (FAPERJ).

\appendix
\section{Oscillator equation with damping term}
\label{app:cqd2}

In Sec.~\ref{sec:eft}, we approximated $\dot{q}\simeq 0$ and used the oscillator equation $c_{\dot{q}^2}\ddot{q} \simeq -c_{(2)}q$ to relate $c_{\dot{q}^2}$ to the $\varphi$-mode frequency as in Eq.~\eqref{eq:cq_special}.
In this Appendix, we investigate the effect of the monopole radiation-reaction force as a damping term.

In Appendix~\ref{app:RR}, we show that the radiation-reaction force for an isolated compact object is given by  $\mathcal{F}_q = -\dot{q}$. Inserting that force on the right-hand side of the oscillator equation \eqref{eq:oscillator} yields
\begin{equation} 
c_{\dot{q}^2} \ddot{q}  + V'(q) = \varphi_0 + \mathcal{F}_q.
\end{equation}
Assuming vanishing cosmological scalar field $\varphi_0 = 0$, and keeping only the leading term in $V'(q)$, we get
\begin{equation} 
c_{\dot{q}^2} \ddot{q} + \dot{q} + c_{(2)} q = 0,
\end{equation}
which is the equation for a damped harmonic oscillator and can be written as
\begin{equation} 
\ddot{q} + \gamma_0 \dot{q} + \omega_0^2 q = 0,
\end{equation}
with the definitions
\begin{equation}
\gamma_0 \equiv \frac{1}{c_{\dot{q}^2}} \,, \qquad 
\omega_0^2 \equiv \frac{c_{(2)}}{c_{\dot{q}^2}} \,.
\end{equation}

The solution of this equation is given by
\begin{equation}
q(t) = e^{-\omega_I t} \left[a \cos(\omega_R t) + b \sin(\omega_R t)\right],
\end{equation} 
where
\begin{equation}
\omega_I \equiv \frac{\gamma_0}{2} \,, \qquad  \omega_R \equiv \sqrt{\omega_0^2 - (\gamma_0/2)^2} \,,
\end{equation}
leading to $\omega_0^2 = \omega_R^2 + \omega_I^2$.
These relations can be inverted to obtain $c_{\dot{q}^2}$ and $c_{(2)}$ in terms of $\omega_R$ and $\omega_I$, i.e.,
\begin{equation}
\label{c2cqd2}
c_{\dot{q}^2} = \frac{1}{2\omega_I}, \qquad
c_{(2)} = \frac{\omega_0^2}{2\omega_I}.
\end{equation}

It was argued in Ref.~\cite{Maggiore:2007nq} that $\omega_R$ and $\omega_I$ can be identified as the real and imaginary parts, respectively, of the mode frequency.
Hence, the relation~\eqref{c2cqd2} for $c_{\dot{q}^2}$, obtained by including a damping term in the oscillator equation, is the same as the one in Eq.~\eqref{eq:cq_special}, which was obtained by assuming $\dot{q}\simeq 0$.
However, $c_{(2)}$ calculated from the $\varphi$-mode frequency in Eq.~\eqref{c2cqd2} gives a different result from the one calculated from the quartic fit in Fig.~\ref{fig:potential}, with a relative difference of about $\sim 50\%$. Such a difference would then propagate to $c_{\dot{q}^2}$, but we showed in Fig.~\ref{fig:params} that the scalarization time has a weak dependence on the value of $c_{\dot{q}^2}$. Thus, we choose to identify  $c_{\dot{q}^2} = c_{(2)} / \omega_0^2$, with $c_{(2)}$ calculated from the quartic fit, which is more accurate.

\section{Scalar fluxes and radiation-reaction force}
\label{app:RR}
In this Appendix, we derive the leading order scalar monopole energy flux, and the scalar dipole energy and angular momentum fluxes, following the steps in Ref.~\cite{Damour:1992we}. From the fluxes, we obtain the radiation-reaction force that enters the equations of motion~\eqref{EOMs}.

\subsection{Energy flux}
The scalar-field equation is given by 
\begin{equation}
\square \varphi = - 4 \pi S,
\end{equation}
with the source term
\begin{equation}
S = q_A(t) \delta^3(\bm{x} - \bm{x}_A) + q_B(t) \delta^3(\bm{x} - \bm{x}_B).
\end{equation}
The scalar field has the solution
\begin{align}
\varphi &= \int d^3x' \, \frac{S(x', t_\text{ret})}{|\bm{x} - \bm{x}'|}\,,
\end{align}
where $\bm{x}$ is the distance to the detector at near spatial infinity, and the retarded time $t_\text{ret} = t - |\bm{x} - \bm{x}'|$. 

Defining $R \equiv |\bm{x}|$, $\bm{N} \equiv \bm{x}/R$, and  $t' \equiv t - R$, we expand the scalar field, in the far zone, in terms of multipole moments such that 
\begin{align}
\varphi &= \int d^3x' \, \left[\frac{S(x', t')}{R}
 + \frac{1}{R} \frac{\partial}{\partial t} S(x', t') \bm{x}\cdot \bm{N} + \dots\right] \nonumber\\
& \equiv \frac{1}{R} \left(\Psi + \dot{\Psi}^i N^i  + \dots\right),
\end{align}
where the monopole and dipole moments are defined by
\begin{align}
\Psi &\equiv \int d^3x \, S = q_A + q_B, \\ 
\Psi^i &\equiv \int d^3x\, x^i S = \left( \frac{m^0_B}{M} q_A - \frac{m^0_A}{M} q_B \right) r^i,
\end{align}
and we have used the center-of-mass relations $\bm{x}_A = m^0_B \bm{r} / M, ~ \bm{x}_B = - m^0_A \bm{r} / M$, and $\bm{r} = \bm{x}_A - \bm{x}_B$.

The energy flux due to scalar radiation is related to the stress-energy tensor via
\begin{equation}
\Phi_E = - R^2 \int d\Omega \, T_{0i} N^i,
\end{equation}
where $T_{0i} = \partial_0\varphi \partial_i\varphi /4\pi \simeq -N_i (\partial_0\varphi)^2 /4\pi$, since $\partial_i\varphi \simeq - N_i \partial_0 \varphi + \mathcal{O}(r/R)$, which yields
\begin{align}
\label{fluxMultipoles}
\Phi_E  &= \frac{1}{4\pi} \int d\Omega \, (\dot{\Psi} + \ddot{\Psi}^i N_i + \dots)^2 \nonumber\\
& = \dot{\Psi}^2  + \frac{1}{3}\ddot{\Psi}^i \ddot{\Psi}^i + \dots.
\end{align}
The first term in this equation is the monopole flux, while the second is the dipole flux. To evaluate the angular integral, we have used the relation $\int d\Omega \, N^iN^j = 4\pi \delta^{ij}/3$. 

In taking the time derivatives, we assume that $\dot{q}$ is small compared to $q$. So we neglect terms with $\dot{q}$ in the dipole part of the energy flux, but we keep them in the leading-order monopole flux.
Hence, we obtain
\begin{equation}
\Phi_E \simeq \left(\dot{q}_A + \dot{q}_B\right)^2 + \frac{1}{3} \left(\frac{m_B^0}{M} q_A - \frac{m_A^0}{M} q_B\right)^2 \ddot{\bm{r}}^2,
\end{equation}
with
\begin{equation}
\ddot{\bm{r}} = - \frac{M}{r^2} \left(1 + \frac{q_A q_B}{M\mu}\right) \hat{\bm{r}},
\end{equation}
leading to the monopole and dipole energy fluxes
\begin{align}
\Phi_E^\text{mon} &= \left(\dot{q}_A + \dot{q}_B\right)^2, \label{monFlux} \\
\Phi_E^\text{dip} &= \frac{M^2}{3r^4} \left(\frac{m_B^0}{M} q_A - \frac{m_A^0}{M} q_B\right)^2  \left(1 + \frac{q_A q_B}{M\mu}\right)^2.  \label{dipFlux} 
\end{align}

The dipole flux vanishes for equal masses and charges, in which case, the next-to-leading order monopole flux and the leading order quadrupole flux become the leading contributions.
The scalar quadrupole flux can be computed following the same steps outlined above, with the additional contribution $\dddot{\Psi}^{ij}\dddot{\Psi}^{ij}/15 $ in Eq.~\eqref{fluxMultipoles}. 
However, it is at the same post-Newtonian order as the next-to-leading monopole flux, which requires the equations of motion at next-to-leading order, and is thus beyond the scope of this work.

\subsection{Angular momentum flux}
The angular momentum carried by the scalar field is related to the stress-energy tensor via
\begin{align}
J^i &= \epsilon^{ikl} \int d^3x\, x^kT^{0l} \nonumber\\
&= \frac{1}{4\pi} \epsilon^{ikl} \int d^3x\,\, (\partial_0 \varphi) x^k (\partial^l \varphi).
\end{align} 
Using $\partial^j N^i = (\delta^{ij} - N^jN^j)/R$, and $d^3x = R^2dtd\Omega$, the above equation yields
\begin{align}
\frac{dJ^i}{dt} &= \frac{1}{4\pi }\epsilon^{ikl} \int d\Omega R N^k
\left[\dot{\Psi} + \ddot{\Psi}^j N^j\right] \nonumber \\
&\quad 
\times \left[- N^l\dot{\Psi} - N^lN^m\ddot{\Psi}^m + \frac{1}{R} \left(\delta^{ml} - N^mN^l\right) \dot{\Psi}^m\right].
\end{align}
Due to the antisymmetry of $\epsilon^{ikl}$, the only term that does not vanish is
\begin{align}
\frac{dJ^i}{dt} &= \frac{1}{4\pi }\epsilon^{ikl} \int d\Omega\, 
N^kN^l \ddot{\Psi}^j\dot{\Psi}^l \nonumber\\
& = \frac{1}{3} \epsilon^{ikl} \ddot{\Psi}^k\dot{\Psi}^l\,.
\end{align}
Differentiating the dipole moment, while assuming $\dot{q}\simeq 0$, leads to 
\begin{equation}
\frac{dJ^i}{dt} = \frac{1}{3}\epsilon^{ikl} \left(\frac{m_B^0}{M} q_A - \frac{m_A^0}{M} q_B\right)^2 \ddot{r}^k\dot{r}^l.
\end{equation}
Thus, we obtain the angular momentum flux
\begin{align}
\Phi_J^\text{dip} &= -\frac{dJ^\theta}{dt} \nonumber \\
&\simeq \frac{1}{3}\frac{ML}{\mu r^3} \left(\frac{m_B^0}{M} q_A - \frac{m_A^0}{M} q_B\right)^2 \left(1 + \frac{q_A q_B}{M\mu}\right).
\end{align}

\subsection{Radiation-reaction force}
To obtain the monopole radiation-reaction force, we use the balance equation $\dot{E}_\text{system}^\text{mon} = -\Phi_E^\text{mon}$, where the energy loss by the system is the time derivative of the Hamiltonian in Eq.~\eqref{Hamiltonian} after substituting the equations of motion~\eqref{EOMs} without the dipole or quadrupole radiation-reaction force, leading to
\begin{equation}
\dot{E}_\text{system}^\text{mon} = \mathcal{F}_{q_A}^\text{mon} \dot{q}_A + \mathcal{F}_{q_B}^\text{mon} \dot{q}_B.
\end{equation}
Setting this equal to the energy flux in Eq.~\eqref{monFlux}, and solving for the radiation-reaction force yields
\begin{equation}
\label{scalarMonRR}
\mathcal{F}_{q_A}^\text{mon} = \mathcal{F}_{q_B}^\text{mon} = - \dot{q}_A - \dot{q}_B.
\end{equation}
This force is the same for both $q_A$ and $q_B$, which implies that the equation of motion for the combination $q_A-q_B$ does not have a damping term at that order.
Therefore, for equal masses but opposite charges, such that the sum $q_A+q_B$ vanishes, the monopole force would not affect the charges.
The damping in this case would be provided by terms in the scalar \emph{dipole} flux with time derivatives of the scalar charges.
We dropped these contributions in Eq.~\eqref{dipFlux} for simplicity, but we checked that they have negligible effect on the results considered in this paper.

For the dipole radiation-reaction forces, we follow the arguments in Ref.~\cite{Bini:2012ji}, and use the energy and angular momentum balance equations including Schott terms, which represent additional contributions to energy and angular momentum due to interaction with the radiation field, i.e.,
\begin{align}
\dot{E}_\text{system}^\text{dip} + \dot{E}_\text{Schott} + \Phi_E^\text{dip} &= 0, \nonumber\\
\dot{J}_\text{system}^\text{dip} + \dot{J}_\text{Schott} + \Phi_J^\text{dip} &= 0.
\end{align}
The energy and angular momentum losses by the system are given by
\begin{align}
\dot{E}_\text{system}^\text{dip} &=\frac{dH}{dt}= \dot{r}\mathcal{F}_r^\text{dip} + \dot{\varphi} \mathcal{F}_\varphi^\text{dip} \nonumber\\ 
\dot{J}_\text{system}^\text{dip} &=\frac{dL}{dt}=\mathcal{F}_\varphi^\text{dip}.
\end{align}

We choose to set $\dot{J}_\text{Schott} = 0$, which corresponds to part of the coordinate gauge freedom. 
The components of the radiation-reaction force are then related to the fluxes via
\begin{align}
\dot{r} \mathcal{F}_r^\text{dip} + \dot{E}_\text{schott} &= - \Phi_{EJ}, \nonumber\\
\mathcal{F}_\varphi^\text{dip} &= - \Phi_J,
\end{align}
where we define $\Phi_{EJ} \equiv \Phi_E - \dot{\phi}\Phi_J$ with $\dot{\phi}$ being the orbital frequency.

For circular orbits $\Phi_{EJ} = 0$ since $\Phi_E = \dot{\phi} \Phi_J$. Hence, $\Phi_{EJ}$ can be written in terms of quantities that vanish for circular orbits, such as $p_r^2$ and $\dot{p}_r$, which also satisfy time-reversal symmetry,
\begin{align}
\Phi_{EJ} &= f_1 p_r^2 + f_2 \dot{p}_r \nonumber\\
&= p_r \left(p_r f_1 - \frac{df_2}{dt}\right) + \frac{d}{dt}(f_2 p_r),
\end{align}
for some arbitrary functions $f_1$ and $f_2$, which leads to
\begin{align}
E_\text{Schott} &= - f_2 p_r, \nonumber\\
\mathcal{F}_r^\text{dip} &= - \frac{p_r}{\dot{r}} \left(p_r f_1 - \frac{df_2}{dt}\right).
\end{align}

Applying that approach using the dipole fluxes from Eq.~\eqref{dipFlux} leads to 
\begin{align}
f_1 &= 0, \nonumber\\
f_2 &= - \frac{M}{3\mu r^2} \left(\frac{m_B^0}{M} q_A - \frac{m_A^0}{M} q_B\right)^2 \left(1 + \frac{q_A q_B}{M\mu}\right).
\end{align}
Thus, we obtain the Schott energy and $\mathcal{F}_r^\text{dip}$
\begin{align}
E_\text{Schott} &= \frac{M}{3\mu r^2}p_r \left(\frac{m_B^0}{M} q_A - \frac{m_A^0}{M} q_B\right)^2  \left(1 + \frac{q_A q_B}{M\mu}\right), \\
\mathcal{F}_r^\text{dip} &= \frac{2}{3} \frac{M}{\mu r^3} p_r \left(\frac{m_B^0}{M} q_A - \frac{m_A^0}{M} q_B\right)^2  \left(1 + \frac{q_A q_B}{M\mu}\right).
\end{align}

\bibliography{dynamical_scalarization}

\end{document}